\documentclass[doublecol]{epl2}

\usepackage{amsmath,amssymb,mathrsfs}
\usepackage{graphicx}
\usepackage{latexsym}

\allowdisplaybreaks[4]

\begin{document}
\title{Efficiency and power of a thermoelectric quantum dot device}
\author{D.M.\ Kennes, D.\ Schuricht and V.\ Meden}
\institute{
Institut f\"ur Theorie der Statistischen Physik and JARA---Fundamentals of Future 
Information Technology, RWTH Aachen University, 52056 Aachen, Germany}

\abstract{We study linear response and nonequilibrium steady-state 
thermoelectric transport through a single-level quantum dot tunnel 
coupled to two reservoirs held at different temperatures as well as chemical 
potentials. A fermion occupying the dot interacts with those in the 
reservoirs by a short-ranged two-particle interaction. For parameters for which
particles flow against a bias voltage from the hot to the cold reservoir this setup
acts as an energy-conversion device with which electrical energy is gained 
out of waste heat. We investigate how 
correlations affect its efficiency and output power. In linear response 
the changes in the thermoelectric properties can be traced back to the interaction 
induced renormalization of the resonance line shape. In 
particular, small to intermediate repulsive interactions reduce the 
maximum efficiency. In nonequilibrium the situation is more complex and we 
identify a parameter regime in which for a fixed lower bound of 
the output power the efficiency increases.}

\pacs{73.50.Lw}{Thermoelectric effects}
\pacs{05.60.Gg}{Quantum transport}
\pacs{73.63.Kv}{Electronic transport in nanoscale materials and structures: QD}

\maketitle


\section{Introduction} 
Quantum dots (QDs) are considered as high-potential solid-state energy 
conversion devices to gain electrical energy out of waste heat \cite{Mahan96}. 
In a minimal setup electrons from a hot right reservoir (at temperature $T_{R}$) 
are transported via a weakly coupled QD to a cold left one (at $T_{L}$) 
against a bias voltage $V=\mu_{L} - \mu_{R} \geq 0$. 
Here $\mu_\alpha$ denotes the chemical potential of reservoir $\alpha=L,R$ and we use 
units in which the elementary charge $e=1$. 
Within \textit{linear response}, that is for small temperature differences and small bias 
voltages, one aims at a high efficiency $\eta$ of energy conversion. 
It is defined as $\eta = P/|I_{\rm h}|$, with the output power $P = V |I_{\rm c}|$ and the charge/heat currents
$I_{\rm c/h}$ \cite{footnote}. As a measure of $\eta$ in QD setups one often uses the so-called figure 
of merit $ZT$ \cite{Mahan96};
the larger $ZT$ the closer $\eta$ comes to its upper bound, the Carnot 
efficiency $\eta_{\rm C}$.
We here neglect the contribution of the phonon thermal conductance to $ZT$ and focus 
on the electronic degrees of freedom. The figure of merit is inversely proportional 
to the Lorenz number. Under the 
assumption of scattering processes which on the scale $T_{R} \approx T_{L} \approx T$ only 
weakly depend on energy the Wiedemann-Franz law states that the latter is a material
independent number. This limits the electronic $ZT$ in bulk materials which obey this 
law. In QDs due to the strong energy dependence of the transmission close 
to transport resonances \cite{Mahan96,Mahan97,Kubala08} the Wiedemann-Franz law 
does not apply as long as the temperature is larger than the resonance width $\Gamma$. 
Therefore, in transport through dots with sharp resonances a large figure of merit and 
thus a high efficiency can be realized \cite{Mahan96,Mahan97,Kubala08,Murphy08}. 
Equivalently, QDs can be used as efficient solid-state systems for cooling of 
nanoelectronic devices \cite{Giazotto06}. 

Due to  the confinement of electrons in QDs to  meso- or nanoscopic regions the local 
Coulomb interaction becomes a relevant energy scale. This two-particle interaction 
strongly alters the line shape of resonances and thus the 
thermoelectric properties \cite{Kubala06,Kubala08,Murphy08}. 
For example, the Kondo effect, resulting from correlated spin fluctuations, 
leads to a very sharp many-body resonance of 
width much smaller than the noninteracting $\Gamma$. This  
has crucial consequences for
thermoelectric transport \cite{Kim03,Scheibner05,Costi10,Andergassen11,Rejec12,Azema12,Roura-Bas12,Hong13}. 
Even in the absence of the Kondo 
effect correlated charge fluctuations affect the line width as was worked out in 
detail for the interacting resonant level model (IRLM) 
\cite{Schlottmann80,Filyov80,Bohr07,Borda07,Karrasch10a,Andergassen11a}.  
For weak to intermediate local Coulomb repulsions the interaction 
renormalized width $\Gamma^{\rm ren}$ is increased compared to the noninteracting 
one $\Gamma$, while for strong 
repulsive interactions as well as attractive ones $\Gamma^{\rm ren} < \Gamma$. Effectively 
attractive interactions might be realized in molecular QDs with strong local electron-phonon 
coupling \cite{Andergassen11}. The consequences of the interaction renormalized resonance line shape 
for the thermoelectric transport properties were not discussed so far. We fill this gap and show 
that the parameter dependence of $\eta$ and $P$ can be 
understood in terms of the line width renormalization. We provide approximate analytical 
results for the efficiency and the output power which for small 
two-particle interactions agree very well with numerical data. For weak repulsive 
interactions the maximum efficiency decreases, it increases for attractive ones. 
Depending on the temperature regime considered the two-particle interaction can lead 
to an increase or to a decrease of the power at maximum efficiency.

In linear response the output power is of order $\Delta T^2$, with $\Delta T = T_{L} - T_{R} <0$, 
and thus small (see below). Depending on the precise technological conditions under which 
the QD `heat engine' is supposed to perform (fixed or tunable $\Delta T$, limited or 
unlimited supply of heat, 
scalability in parallel or series, etc.) it might be meaningful to maximize $P$ instead of 
$\eta$ by varying the model parameters and the voltage at fixed temperatures, 
with $|\Delta T|$ not necessarily being small. Within the framework of nonequilibrium 
thermodynamics studying the efficiency at maximum power of general heat 
engines is an active field of current 
research \cite{Curzon74,VandenBroeck05,Schmiedl08,Esposito10a,Benenti11,Yan12,Brandner12}. In this 
context the noninteracting resonant level model in the limit of vanishing level-reservoir coupling was studied as a 
toy model \cite{Esposito09}. Even questions
such as `what is the optimal efficiency reachable for the power being larger than a given lower 
bound' might be of interest. Computing the efficiency at maximum power and answering questions of the above type 
requires access to the full \textit{nonequilibrium} (in $V$ and $\Delta T$) \textit{steady-state}
properties \cite{Leijnse10}. For microscopic models of QDs with local two-particle correlations achieving this 
presents a formidable challenge; for a recent review, see e.g.~Ref.~\cite{Andergassen10}. 
In particular, this holds if nonperturbative effects in both
the level-reservoir coupling as well as the two-particle interaction become crucial as it is the case in 
the IRLM \cite{Schlottmann80,Filyov80,Bohr07,Borda07,Karrasch10a,Andergassen11a}. 

Only recently a very flexible 
nonperturbative tool was developed which allows to treat the full-fledged nonequilibrium steady state 
of the IRLM \cite{Karrasch10a,Karrasch10b,Kennes12} at weak interactions. It is based on 
Keldysh Green functions \cite{Rammer07} and the 
functional renormalization group approach to quantum many-body physics \cite{Metzner12}. 
Here we apply this method to study the efficiency and power output of the IRLM `heat engine' 
beyond linear response. Broadly speaking the effect of the interaction on the efficiency (at 
maximum power) is similar to the one found in linear response. By closer inspection we identify a situation in which 
a weak repulsive interaction can lead to an increase of the efficiency by a few percent 
at fairly large power output.    
  
We here focus on correlation effects in the elementary two-reservoir IRLM 
but expect our results to be of relevance 
for other models showing correlated charge fluctuations as well.  
Very recently three-terminal systems with the reservoirs having more complex 
degrees of freedom (e.g. magnetic ones) were identified as 
promising energy converters \cite{Entin-Wohlman10,Sanchez11,Sothmann12}. In those systems 
electrons are transported against a bias between two of the reservoirs by extracting heat from the third one. 

\section{Quantum dot model}  
\begin{figure}[t]
\centering
\includegraphics[width=0.5\linewidth,clip]{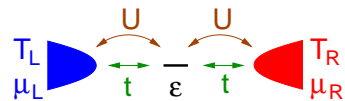}
\caption{Sketch of the investigated model.}
\label{figmodel}
\end{figure}

Our model Hamiltonian sketched in Fig.~\ref{figmodel} consists of three parts
$H =H_{\rm r} + H_{\rm d} + H_{\rm c}$. The two spinless reservoirs are given by 
\begin{equation}
H_{\rm r }=\sum_{k ,\alpha} \epsilon_k \, c_{k,\alpha}^\dag c_{k,\alpha}^{} , 
\label{leadham}
\end{equation}
with fermionic ladder operators $c_{k,\alpha}^{(\dag)}$, where 
$k$ denotes a set of quantum numbers characterizing the reservoir states. To disentangle 
effects of the reservoir band structure  encoded in the dispersion $\epsilon_k$ and the correlations 
on the thermoelectric properties we assume structureless reservoirs with an energy independent 
density of states $\rho$ between $-D$ and $D$ and the bandwidth $2D$ being much larger than
any other energy scale. 
The reservoirs are in grand canonical equilibrium 
with $T_\alpha=1/\beta_\alpha$ and chemical potentials centered around zero: 
$\mu_L=-\mu_R=V/2$ (with $\hbar=k_{\rm B}=e=1$). 
The dot part of $H$ is given by $H_{\rm d}= \epsilon n$, with 
the dot occupation number operator $n = d^\dag d^{}$. The energy $\epsilon$ 
of the level is assumed to be tunable; in experiments this is achieved by 
applying a voltage to a properly fabricated gate. Finally, the coupling reads 
\begin{equation}
 H_{\rm c} = t \! \sum_{k ,\alpha} \! \left( c^\dag_{k,\alpha} d + \mbox{H.c.} \right)
+ u  \left( n - \frac{1}{2} \right) \! \sum_{k,k',\alpha} \!\!\!  : \!  c_{k,\alpha}^\dag c_{k',\alpha}^{}  \!\! : ,  
\label{coupham}
\end{equation}
where $: \ldots :$ denotes normal ordering. This and the shift of the dot occupancy by $-1/2$ ensures 
that $\epsilon=0$ corresponds to half dot filling. For simplicity we assume the tunnel couplings 
to the left and right reservoir $t$ to be equal; similarly for the two-particle interactions $u$ 
to the left and right. We note that 
within our approach to the many-body problem these restrictions can easily be relaxed. We 
use the dimensionless interaction $U=\rho u$.

\section{Thermoelectric transport at $U=0$} 
\begin{figure}[t]
  \centering
\vspace{-.3cm}
  \includegraphics[width=0.8\linewidth,clip]{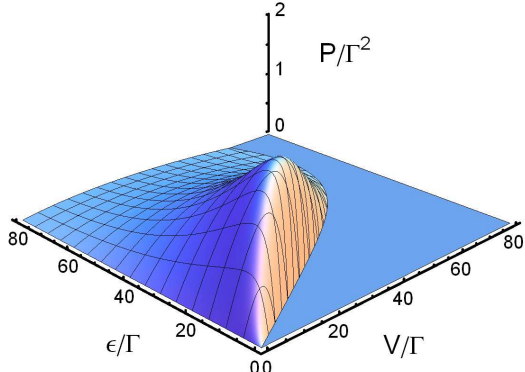}
\vspace{-.5cm}
  \caption{(Color online) Output power $P$ of our quantum dot device for vanishing 
two-particle interaction as a function of $\epsilon/\Gamma$ and $V/\Gamma$ for 
$T_L/\Gamma=1$ and $T_R/\Gamma=20$. Data points are only shown in the regime in which the device 
 acts as a `heat engine' (curved surface with lines).} 
  \label{figpepsV}
\end{figure}

For $U=0$ transport through the dot is characterized by a Breit-Wigner transmission resonance 
\begin{equation}
\tau(\omega) = \frac{\Gamma^2}{(\omega - \epsilon)^2 + \Gamma^2}
\label{reso}
\end{equation}
of width $\Gamma = 2 \pi \rho t^2$. The steady-state charge current $I_{\rm c}$ (leaving the left reservoir) and heat current
$I_{\rm h}$ (entering the right reservoir) can be computed using the Landauer-B\"uttiker formalism
\begin{align}
I_{\rm c} & = \frac{1}{2 \pi} \int d \omega \; \tau(\omega) \left[f_L(\omega) - f_R(\omega)  \right] , \label{eq:IC}\\
I_{\rm h} & = \frac{1}{2 \pi} \int d \omega \; (\omega - \mu_R) \tau(\omega) \left[f_L(\omega) - f_R(\omega)  \right] ,
\label{currents}
\end{align}
with the Fermi functions $f_\alpha (\omega) = \left[ e^{(\omega - \mu_\alpha)/T_\alpha} + 1 \right]^{-1}$. 
We express the energy integral 
in $I_{\rm c}$ in terms of the digamma function $\Psi(0,z)$ as 
\begin{align}
I_{\rm c} (\epsilon) =  \frac{\Gamma}{2\pi} \, \mbox{Im} & \left[ \Psi \left(0,\frac{1}{2} + \frac{i}{2 \pi T_R}[\epsilon - \mu_R - i \Gamma] \right)  
\right. \nonumber \\ &  \left. - \Psi \left(0,\frac{1}{2} + \frac{i}{2 \pi T_L}[\epsilon - \mu_L - i \Gamma] \right)  
\right] 
\label{pcurrentexpl}
\end{align}
and the heat current as a sum of the charge current and 
the Hilbert transform of the latter
\begin{equation}
I_{\rm h} (\epsilon) = (\epsilon - \mu_R) I_{\rm c} (\epsilon) -   
\frac{\Gamma}{\pi} \int \!\!\!\!\!\! {\mathcal P} d \epsilon' \frac{I_{\rm c}(\epsilon')}{\epsilon- \epsilon'} .
\label{ecurrentexpl}
\end{equation}
In Eqs.~(\ref{pcurrentexpl}) and (\ref{ecurrentexpl}) 
$\Gamma$ can be scaled out (taken as the unit of energy) and  $\eta$ 
as well as $P$ can be investigated as functions of the model parameter $\epsilon/\Gamma$, the
external bias voltage $V/\Gamma$, and $T_{L/R}/\Gamma$. 
In analogy to studies 
on periodic heat engines in thermodynamics we assume that $T_{L/R}$ are fixed by the environment.  
With $|\Delta T|/\Gamma$ not necessarily being small we cannot resile to the linear response regime. 
Figure \ref{figpepsV} shows $P=V |I_{\rm c}|$ in the $\epsilon$-$V$-plane for 
$T_L/\Gamma=1$ and $T_R/\Gamma=20$. It has a unique maximum. 

To set the stage for more general considerations we first consider the limit 
$\Gamma \to 0$ \cite{Esposito09}. The second term in Eq.~(\ref{ecurrentexpl}) is of higher order 
in $\Gamma$ as compared to the first one and can be neglected. 
As a consequence charge and heat flow are  `perfectly coupled' 
(proportional to each other); for $\Gamma \to 0$ the energy of every particle 
coming from the dot level becomes sharp and is fixed at $\epsilon$ \cite{Esposito09}.
For applications it is meaningful to maximize the output power---the maximum is denoted 
by $P_{\rm m}$ in the following---and study the efficiency at maximum power $\eta_{\rm mp}$ as a 
function of $T_{L/R}$. 
The optimization is performed with respect to the level energy $\epsilon$ and 
the externally applied voltage $V$ both being parameters which in experiments 
on QDs can routinely be varied with high precision. As shown analytically in 
Ref.~\cite{Esposito09} for $\Gamma \to 0$, $\eta_{\rm mp}$ and $P_{\rm m}$ are 
functions of $T_L/T_R$ only and can thus be written as functions of the 
Carnot efficiency $\eta_{\rm C} = 1-T_L/T_R= |\Delta T|/T_R$. 

\begin{figure}[t]
  \centering
\vspace{-.3cm}
  \includegraphics[width=0.365\linewidth,clip]{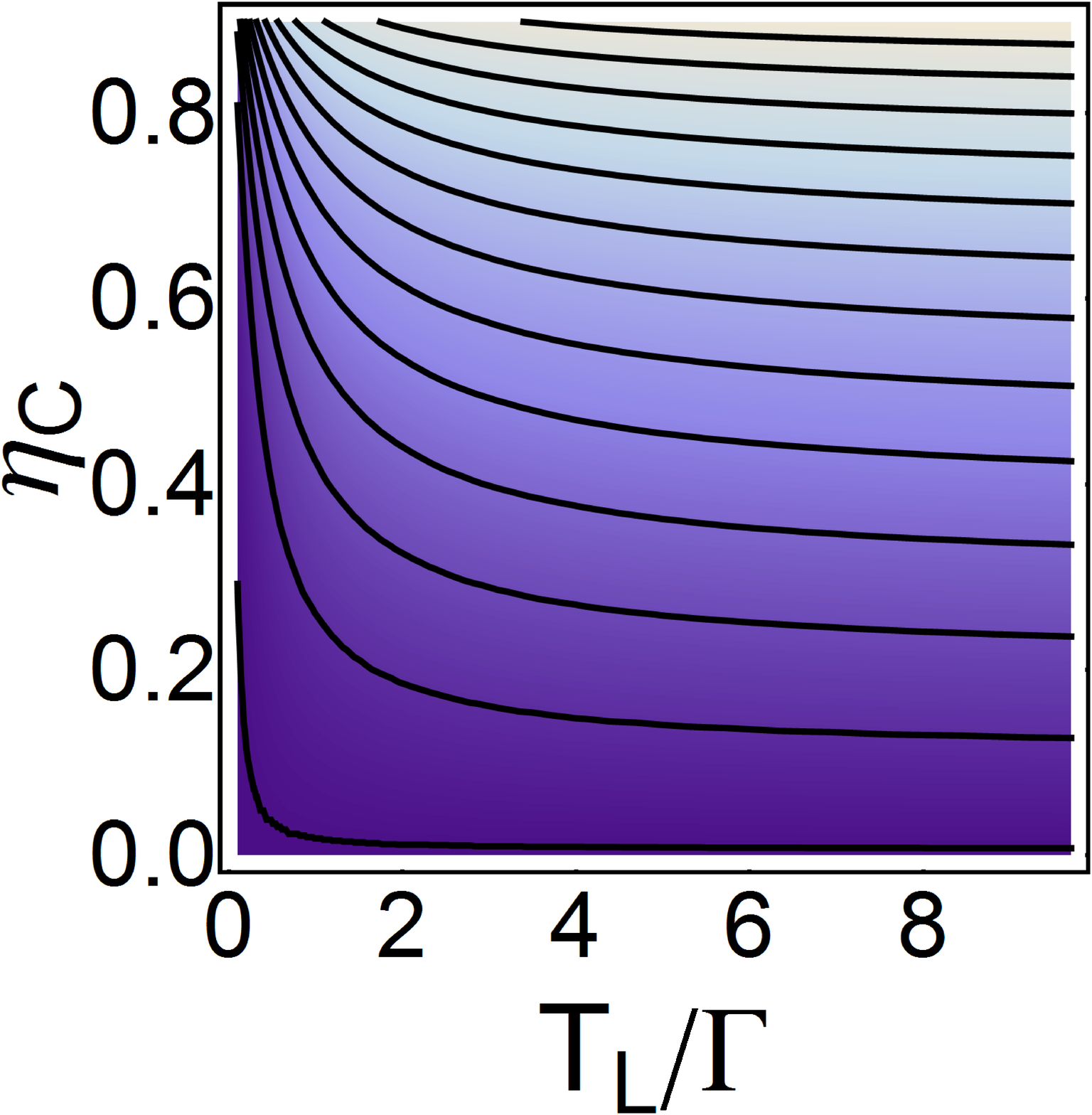} 
  \includegraphics[width=0.113\linewidth]{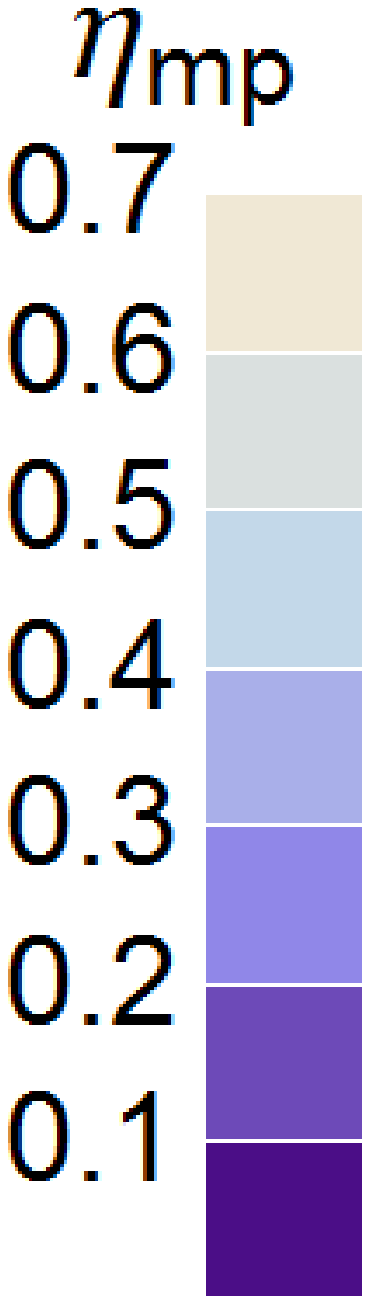}  
\includegraphics[width=0.365\linewidth,clip]{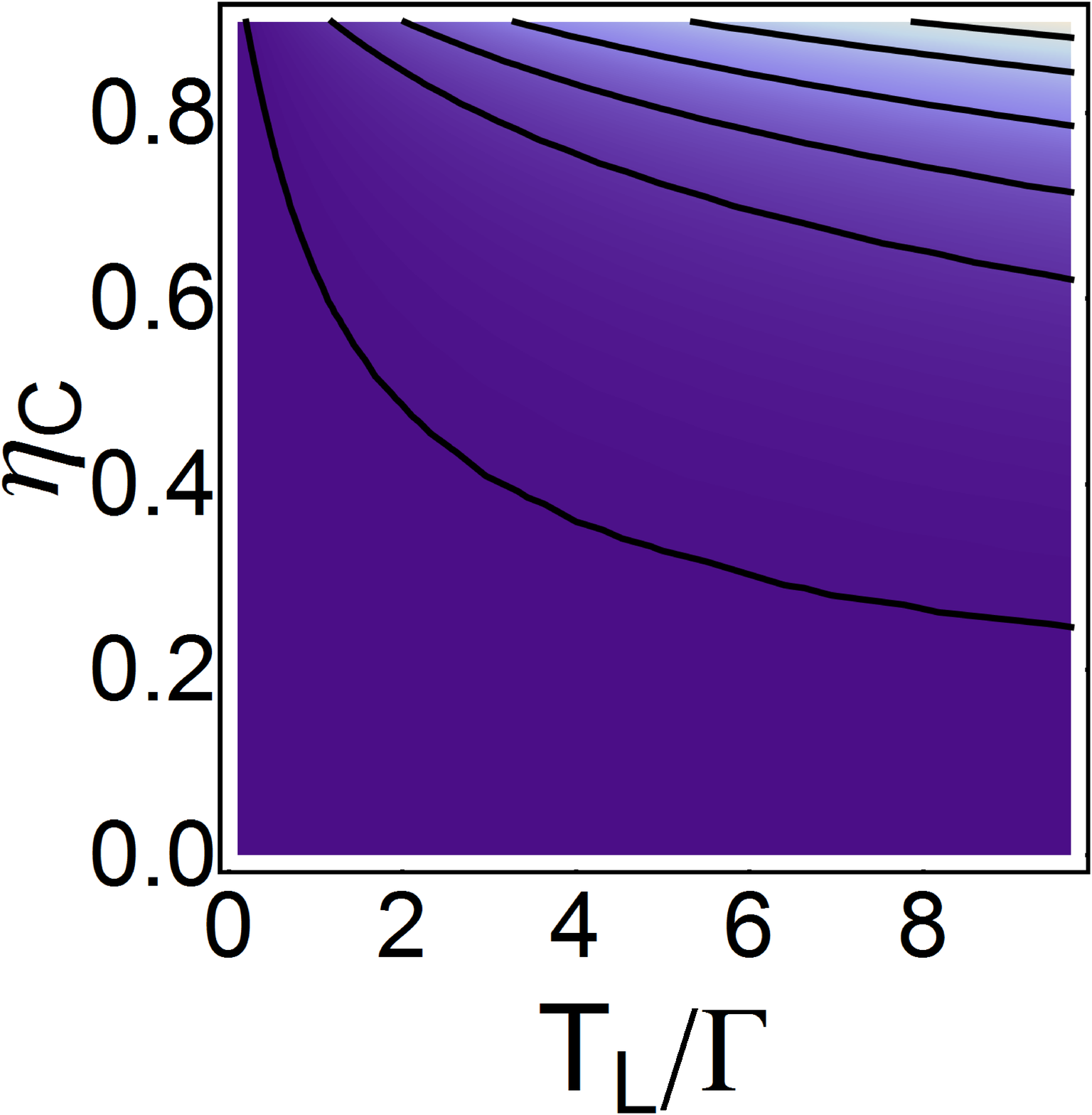}
\includegraphics[width=0.113\linewidth]{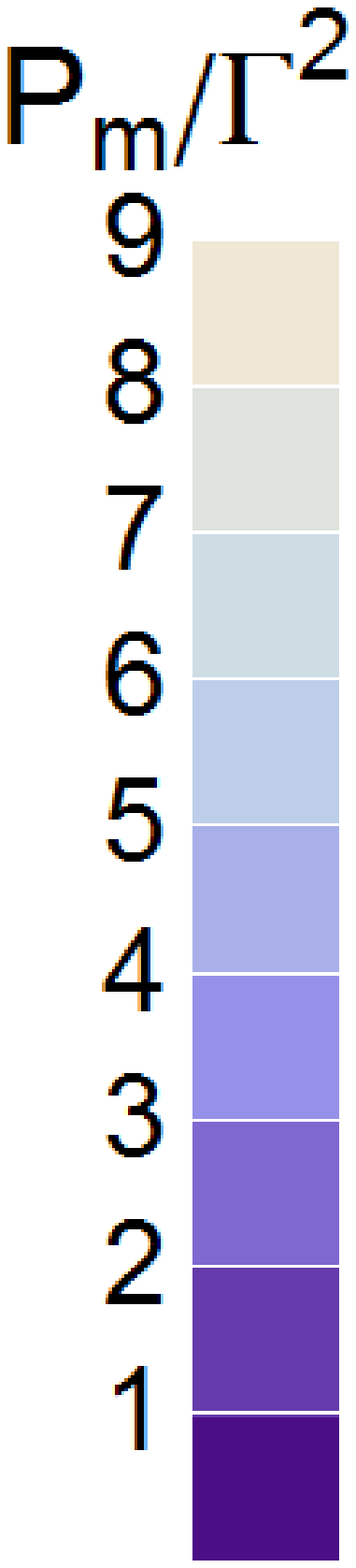}
  \caption{(Color online) The efficiency at maximum power (left) and the
maximum power (right) 
for $U=0$ as functions of $T_L/\Gamma$ and $\eta_{\rm C} = 1-T_L/T_R$. Contour lines are shown.}
  \label{figetacmPm}
\end{figure}

We now return to $\Gamma > 0$ and numerically maximize the output power with 
respect to $\epsilon$ and  
$V$ using Eqs.~(\ref{pcurrentexpl}) and (\ref{ecurrentexpl}). Both,  
$\eta_{\rm mp}$ and $P_{\rm m}$ are no longer functions 
of $\eta_{\rm C}$ only. This becomes apparent from Fig.~\ref{figetacmPm} which shows 
$\eta_{\rm mp}$ and $P_{\rm m}$ as functions of $T_L/\Gamma$ and $\eta_{\rm C}$.
Only in the limit of large $T_L/\Gamma$, that is small $\Gamma$, the dependence 
on $T_L/\Gamma$ drops out. For small $T_L/\Gamma$ rather large $\eta_{\rm C}=|\Delta T|/T_R$ (close to 1) 
are required to obtain a sizable efficiency at maximum power. 

Expanding Eqs.~(\ref{pcurrentexpl}) and 
(\ref{ecurrentexpl}) to lowest order in $\Delta T/\Gamma$ and $V/\Gamma$ 
we obtain linear response results \cite{Murphy08}. The currents 
follow from the Onsager matrix containing the conductances  
\begin{eqnarray} 
\left( \begin{array}{c} I_{\rm c} \\ I_{\rm h} \end{array} \right) = \left(  \begin{array}{cc} G_{\rm c} & G_{\rm c}^{\Delta T} \\ 
G_{\rm h}^{V} & G_{\rm h} \end{array} \right) \left( \begin{array}{c} V  \\ \Delta T \end{array} \right) .
\label{linrespmatrix}
\end{eqnarray}
The Onsager (time-reversal) symmetry gives $G_{\rm h}^{V} =  T G_{\rm c}^{\Delta T}$, with 
$T=T_L = T_R$, and the independent matrix elements can be written as \cite{Murphy08}  
\begin{align}
G_{\rm c}&=\frac{\Gamma}{8 \pi^2 T}
\left[\Psi\left(1,\frac{\pi+w}{2\pi}\right)+ \mbox{c.c.} \right] , \label{trigamma0} \\
  G_{\rm c}^{\Delta T}&=- \frac{i \Gamma}{8 \pi^2 T }
\left[w\Psi\left(1,\frac{\pi+w}{2\pi}\right)- \mbox{c.c.} \right] , \\
  G_{\rm h}&=\frac{ \Gamma}{2\pi  }\left( \frac{\Gamma}{T}- \frac{1}{4\pi}
\left[w^2\Psi\left(1,\frac{\pi+w}{2\pi}\right)
+   \mbox{c.c.} \right]\right) ,
\label{trigamma}
\end{align}  
with the trigamma function $\Psi(1,z)$ and $w=(\Gamma+ i \epsilon)/T$.  

In linear response one aims at a large $\eta$. Independent of the model 
considered one first maximizes $\eta=V |I_{\rm c}|/|I_{\rm h}|$ with respect to $V$
\cite{VandenBroeck05,Benenti11} using Eq.~(\ref{linrespmatrix}) leading to 
\begin{equation}
\eta_0 =    \frac{\sqrt{ZT+1}-1}{\sqrt{ZT+1}+1}   \, \frac{|\Delta T|}{T} , \;\;
ZT = \frac{(G_{\rm c}^{\Delta T})^2 T}{ G_{\rm c} G_{\rm h} - (G_{\rm c}^{\Delta T})^2 T }  .
\label{linrespeta}
\end{equation}  
The maximum is reached for a voltage of order $\Delta T$ and the linear response 
regime is not left.   
Equation (\ref{linrespeta}) constitutes the relation between the efficiency and the 
figure of merit 
$ZT$ announced in the introduction; the latter being expressible in terms of the
conductances. We can now be more precise: $ZT$ is a measure for the linear 
response efficiency \textit{maximized with respect to $V$.} For 
$ZT \to \infty$, $\eta_0$ approaches 
the linear response Carnot efficiency $|\Delta T|/T$ which constitutes an upper bound for 
all thermodynamic efficiencies. The power corresponding to $\eta_0$ is given by 
\begin{equation}
P_0 =   \frac{ T G_{\rm h}}{\sqrt{ZT+1}} \,
\frac{\sqrt{ZT+1}-1}{\sqrt{ZT+1}+1}   \, \left(\frac{\Delta T}{T}\right)^2 
\label{linrespP}
\end{equation}  
and thus of order $(\Delta T)^2$; one factor $\Delta T$ comes from the voltage 
at which the maximum efficiency is reached and the other from the charge current.
In the limit $ZT \to \infty$, with $\eta_0 \to \eta_{\rm C}$, the prefactor of $P_0$ 
vanishes. 
  
We now return to our specific model and similar to the nonequilibrium case above first 
consider the limit $\Gamma \to 0$. Independent of $\epsilon \neq 0$ one finds 
$ZT \sim 1/\Gamma$ \cite{Murphy08} and thus $\eta_0 \to \eta_{\rm C}$ \cite{Esposito09}. 
Furthermore, $ G_{\rm h} \sim \Gamma$ and $P_0 \to 0$. 

\begin{figure}[t]
  \centering
\vspace{-.5cm}
  \includegraphics[width=.685\linewidth,clip]{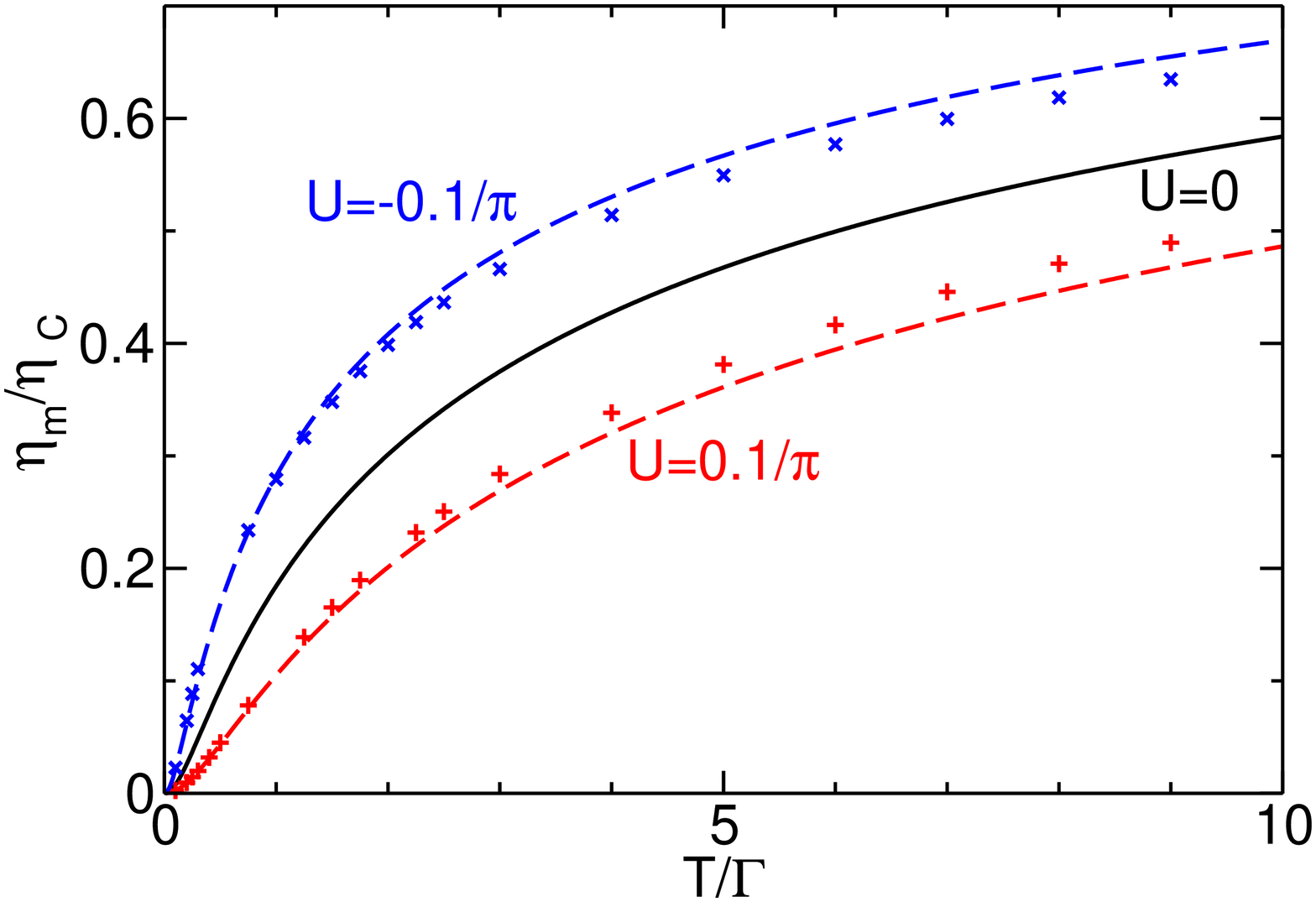}\\
\vspace{-.2cm}
\includegraphics[width=.685\linewidth,clip]{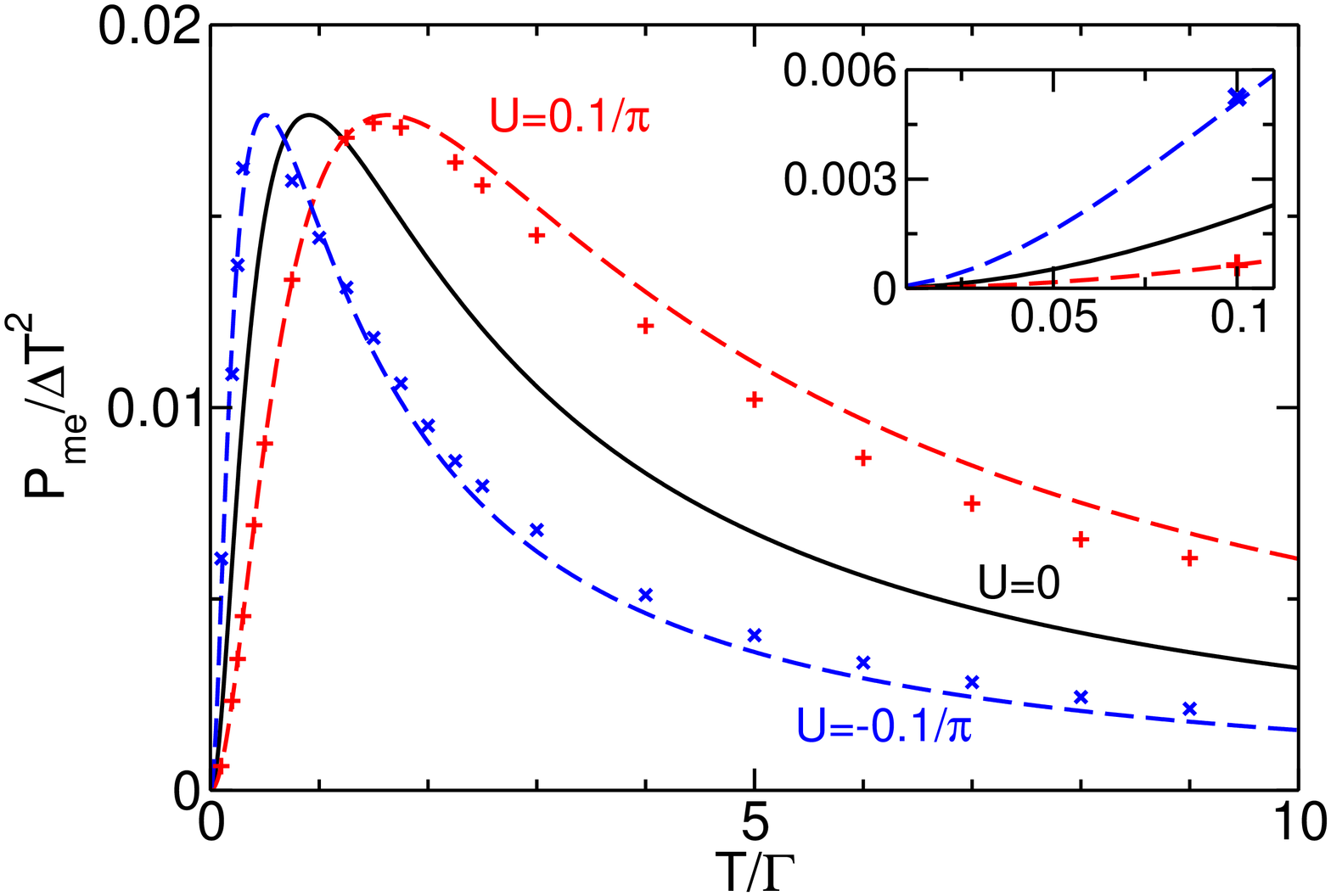}
  \caption{(Color online) Linear response maximum efficiency (upper panel) and the 
corresponding output power (lower panel) 
 for different interactions $U$ as functions of $T/\Gamma$. The lines follow from  
Eqs.~(\ref{linrespmatrix})-(\ref{trigamma}) and (\ref{renwidth}) (for $U \neq 0$) while the symbols 
are numerical data obtained from our many-body approach. Inset: zoom-in of the regime 
$T/\Gamma \ll 1$.}
  \label{figetam}
\end{figure}

For $\Gamma >0$ we numerically maximize $\eta_0$ (in addition) with respect to 
$\epsilon$ using Eqs.~(\ref{linrespmatrix})-(\ref{trigamma}) \cite{Murphy08}.
The resulting efficiency $\eta_{\rm m}$ and power at maximum 
efficiency $P_{\rm me}$ 
as functions of $T/\Gamma$ are shown in Fig.~\ref{figetam}. 
As discussed above for $\Gamma \to 0$, that is for large $T/\Gamma$, $\eta_{\rm m}$ 
approaches Carnot efficiency and $P_{\rm me}/\Delta T^2$ vanishes.  
The latter also vanishes for $T/\Gamma \to 0$ and is thus a 
nonmonotonic function with a maximum at $T \approx \Gamma$. 
Our calculations show that for small $T$  the level energy at which optimal efficiency is reached 
approaches a constant. Using this and the expansions of 
Eqs.~(\ref{trigamma0})-(\ref{trigamma}) we obtain $T G_{\rm h} \sim T^2$ and $ZT \sim T^2$.
Thus $P_{\rm me}/\Delta T^2 \sim T^2$ (see the inset of Fig.~\ref{figetam}).  
  
\section{Thermoelectric  transport at $U \neq 0$} 
To investigate the linear response and nonequilibrium steady-state transport properties 
in the presence of two-particle correlations we use a technique which is based on nonequilibrium 
Keldysh Green functions \cite{Rammer07} and the functional renormalization group approach 
to quantum many-body physics \cite{Metzner12}. In this one derives an exact infinite hierarchy 
of coupled differential equations for the many-body self-energy and the effective $n$-particle 
interactions (one-particle irreducible vertex functions). The derivative is taken with respect 
to a low-energy cutoff and the set of equations have to be integrated from infinity to zero. 
Concrete calculations require truncations. We apply the lowest-order approximation in which 
only the self-energy flow is kept. It was earlier used to study the IRLM for $|U| \ll 1$ in 
equilibrium as well as the charge current in steady-state 
nonequilibrium \cite{Karrasch10a,Karrasch10b,Kennes12}. We refrain from giving any technical
details and instead refer the interested reader to Refs.~\cite{Karrasch10a} and \cite{Kennes12}. 

As its main effect the two-particle interaction generates a renormalization group flow of 
the hybridization $\Gamma/2$ to the 
individual reservoirs. For the left one it is cut off by the largest of the energy scales $|\epsilon - \mu_L|$, 
$T_L$, and the total hybridization; for the 
right one one has to replace the index $L$ by $R$ \cite{Karrasch10a,Karrasch10b,Kennes12}. 
This leads to a power-law scaling  $\Gamma_\alpha^{\rm ren}/\Gamma \sim (s/D)^{-\nu(U)} $, 
with the corresponding largest scale $s$ (still much smaller than $D$) and the 
$U$-dependent exponent $\nu$. To leading order one finds $\nu(U) = 2U$ 
\cite{Schlottmann80,Borda07,Karrasch10a,Andergassen11a,Karrasch10b,Kennes12,Doyon07}. 
This correlation effect cannot be captured by perturbative 
(in either $U$ or $\Gamma$) approaches. In a general nonequilibrium setup 
$\Gamma_L^{\rm ren} \neq  \Gamma_R^{\rm ren}$ even in the case of equal bare hybridizations. 
To exemplify how the renormalization manifests in physical observables
we consider the linear response ($V, \Delta T \to 0$) charge conductance $G_{\rm c}$ as a function 
of $\epsilon$ close to the resonance (at $\epsilon=0$) and $T=0$ \cite{Bohr07,Karrasch10a,Andergassen11a}. 
The largest scale cutting off the renormalization group flow of the hybridizations is given by 
$\Gamma$ . Using the above scaling law one concludes that the resonance in 
$G_{\rm c}$ has the width 
\begin{equation}
\Gamma^{\rm ren} = W = \Gamma \, \left( \frac{2 \Gamma}{D} \right)^{-2 U} ,
\label{renwidth}
\end{equation}
which defines the emergent energy scale $W$. The resonance becomes 
wider for weak (to intermediate \cite{Bohr07}) repulsive interactions but more narrow for 
attractive ones. To illustrate how $\Gamma^{\rm ren}$ changes when two or more of the energy 
scales are of the same order \cite{Karrasch10a,Karrasch10b,Kennes12} we consider 
$T \gtrapprox W$. In this case $\Gamma^{\rm ren} \approx W (T/W)^{-2 U}$; see Fig.~1 of Ref.~\cite{Kennes12}. 
Since $|U| \ll 1$ the correction to $W$ is of order one as long as $T$ does not exceed $W$ by several 
orders of magnitude. Further down this will become crucial. 

We start the discussion of our results of correlation effects on the efficiency and power 
considering linear response. The symbols in Fig.~\ref{figetam} show the 
maximum efficiency  $\eta_{\rm m}$ (maximized with respect to $V$ and $\epsilon$) 
and the power at maximum efficiency $P_{\rm me}$ for $U= \pm 0.1/\pi$ obtained by numerically
integrating the renormalization group equations and computing the charge and thermal 
currents \cite{footnoteLB}. For weak repulsive interactions $\eta_{\rm m}$ is reduced
compared to the noninteracting result, while it is increased for weak attractive 
ones. Depending on the temperature regime considered the interaction enhances or reduces $P_{\rm me}$. 
These results can be understood analytically by employing the above discussed 
energy scale dependence renormalization of the resonance width. 
For $U=0$ the level energy $\epsilon$ at which the maximum efficiency is reached is of the order 
of a few $T$ at large $T$ \cite{Murphy08} while it saturates at a constant of order $\Gamma$ 
for small $T$ . This also holds for small $|U|$. 
Thus Eq.~(\ref{renwidth}) gives a very good estimate of the renormalized hybridization 
for all temperatures shown in Fig.~\ref{figetam}. This motivates us to replace $\Gamma$  
in the noninteracting expressions Eqs.~(\ref{linrespmatrix})-(\ref{trigamma}) for the charge 
and heat currents by $W$ Eq.~(\ref{renwidth}). The corresponding results for  $\eta_{\rm m}$ and $P_{\rm me}$ are 
shown as dashed lines in Fig.~\ref{figetam}. As expected they show excellent agreement 
with the numerical data. The deviations at $T/\Gamma \gtrapprox 1$ are consistent with
the finite $T$ correction to $W$ discussed above. 

As our approximate approach is restricted to $|U| \ll 1$ we cannot access the regime 
of large interactions. It is known that $\Gamma^{\rm ren}$ is a nonmonotonic function 
of $U>0$ and for large repulsive interactions one finds $\Gamma^{\rm ren} < 
\Gamma$ \cite{Bohr07,Borda07}. 
This follows from a second order term $\sim - U^2$ 
which dominates for $U >1$ and changes 
the sign of the exponent $\nu$ \cite{Borda07}. Based on our above results it is thus reasonable to 
assume that for large repulsive interactions $\eta$ will be larger than in the noninteracting 
case. It would be very interesting to explicitely verify this using a method complementary to ours.

\begin{figure}[t]
  \centering
\vspace{-.5cm}
  \includegraphics[width=.685\linewidth,clip]{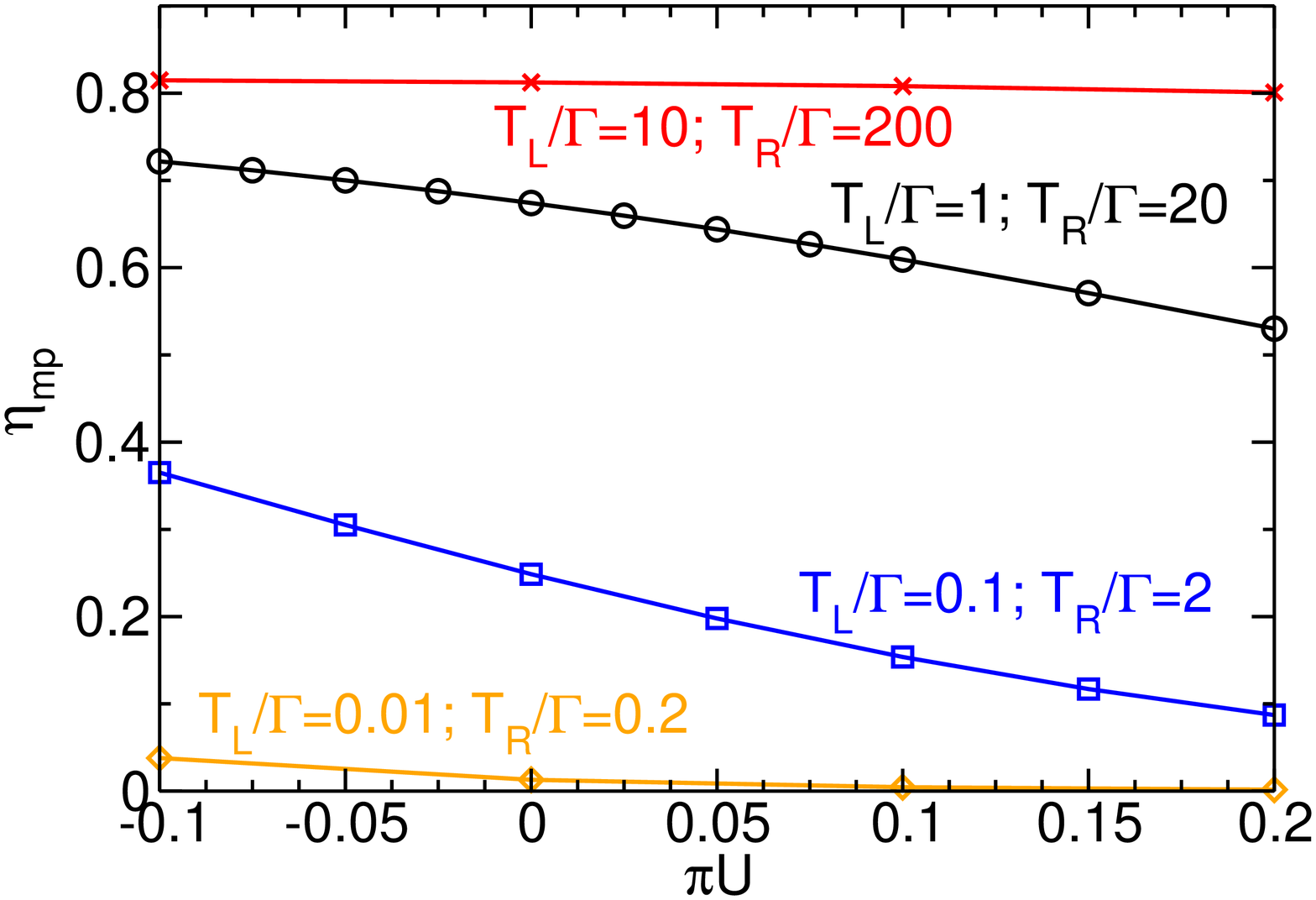} \\
\vspace{-.2cm}
\includegraphics[width=.685\linewidth,clip]{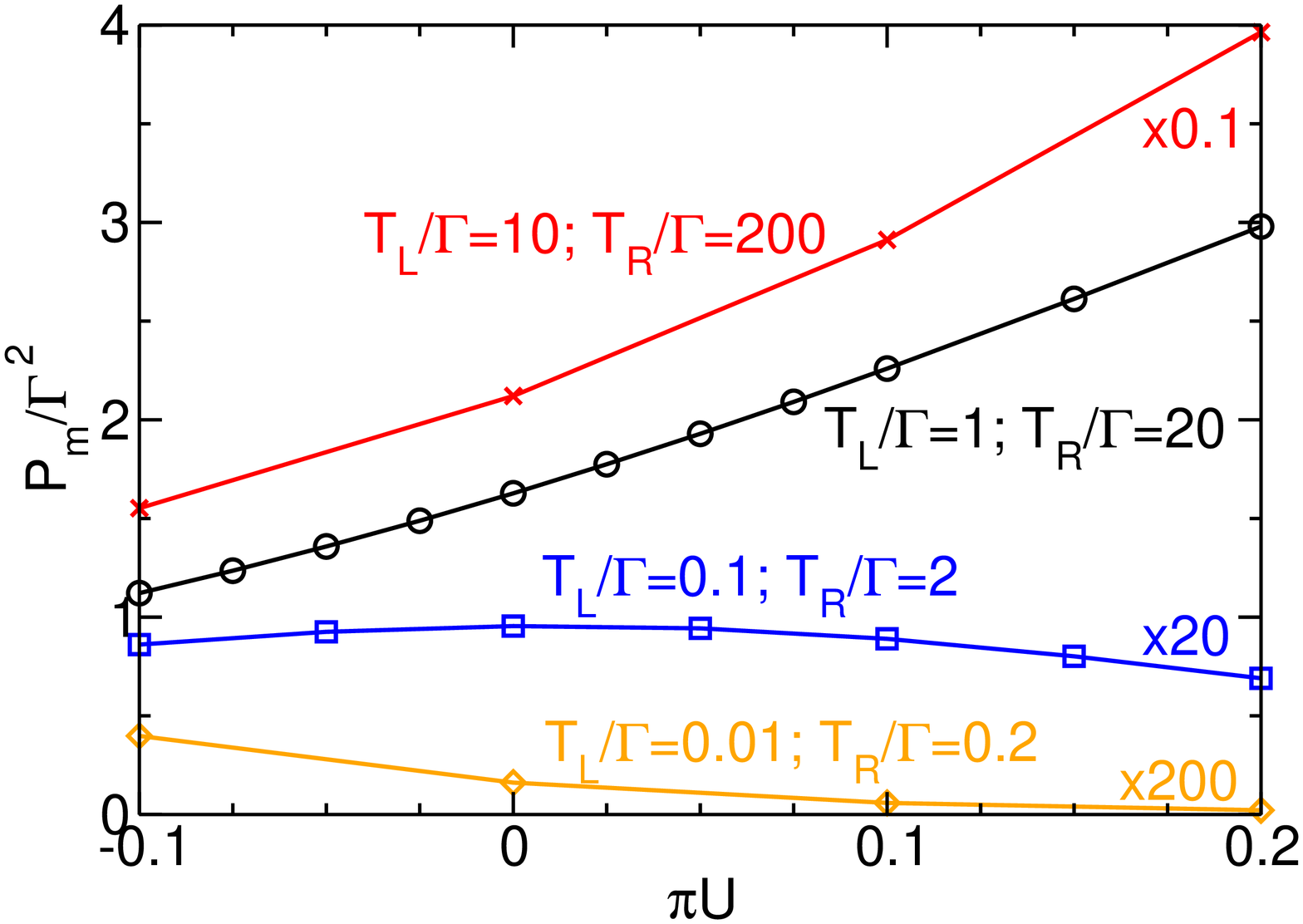}
  \caption{(Color online) Efficiency at maximum power (upper panel) and the 
  maximum power (lower panel) for different sets of $T_{L/R}$ as functions of $U$. 
  The lines are guide to the eyes. The data are scaled by the given factors.}
  \label{maxpowerU}
\end{figure}

We next leave the linear response regime. Figure \ref{maxpowerU} shows the 
$U$ dependence of the maximum power $P_{\rm m}$ and the efficiency at maximum power 
$\eta_{\rm mp}$ for different sets of $T_{L/R}$ obtained 
numerically using our nonequilibrium many-body method. In analogy to the maximum efficiency 
in linear response $\eta_{\rm mp}$ is a decreasing function of $U$. A similar analogy 
to the linear response results for $P_{\rm me}$ Fig.~\ref{figetam} holds for  
the maximum power $P_{\rm m}$.  It is a decreasing function of $U$  
for temperatures $T_{L/R}$ smaller than $\Gamma$ and an increasing one for $T_{L/R}/\Gamma \gg 1$. 
For intermediate temperatures we find nonmonotonic behavior. We have also studied the $U$ dependences
of the maximum (with respect to $V$ and $\epsilon$) efficiency and the corresponding output 
power beyond the linear response regime. 
They are qualitatively similar to the results for $\eta_{\rm mp}$ and $P_{\rm m}$ shown in 
Fig.~\ref{maxpowerU}.

\begin{figure}[t]
  \centering
  \includegraphics[width=0.65\linewidth,clip]{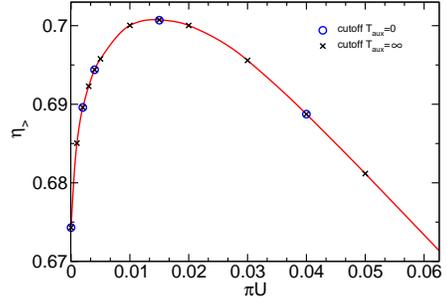}
  \caption{(Color online) Interaction $U$ dependence of the largest efficiency 
   obtainable with $P \geq P_{\rm b}$ for $T_L /\Gamma=1$ and $T_R /\Gamma=20$ 
(for details see the text). The line is an Akima spline interpolating between 
the computed data points (crosses: $T_{\rm aux}=\infty$; circles: $T_{\rm aux}=0$). 
The efficiency can be raised by a few percent by turning on a small repulsive interaction. }
  \label{eta>}
\end{figure}

We finally show that under special conditions the decrease of the efficiency for 
repulsive interactions  can be overcome. To find those
we exploit that for appropriate temperatures $T_{L/R}$ the maximum power 
increases with $U$; this already sets the first requirement,
namely on $T_{L/R}$. We next assume that our `heat engine' has to produce 
at least a certain minimal output power which is a sensible constraint  
if we are interested in charging a battery. For fixed $T_{L/R}$ we proceed as 
follows: as our lower bound of the power $P_{\rm b}$ we for simplicity take 
the maximum power at $U=0$. Varying $\epsilon$ and $V$ we then search for 
the largest efficiency $\eta_{>}$ at fixed $U>0$ with $P \geq P_{\rm b}$. As this is a 
numerically demanding procedure---remind that for every parameter set the currents 
$I_{\rm c/h}$ have to be computed numerically solving coupled differential flow 
equations---we restrict ourselves to one of the temperature sets  
of Fig.~\ref{maxpowerU} namely $T_L /\Gamma=1$ and $T_R /\Gamma=20$. The $U$ dependence of 
$\eta_{>}$ is shown in Fig.~\ref{eta>}. Under the above conditions the efficiency 
increases with increasing interaction even at 
small $U$.  At larger $U$ the generic decrease 
of the efficiency with increasing $U$ takes over leading to a maximum 
at $U \approx 0.005$. It is known \cite{Karrasch10a,Karrasch10b} that using our 
approximate renormalization group based many-body method one achives quantitative 
agreement with numerically exact results \cite{Bohr07} for charge transport of 
the IRLM  up to significantly larger $U$. To demonstrate the robustness of the 
subtle efficiency gain which in addition involves heat transport we considered 
two different cutoff procedures: the reservoir cutoff with 
temperature $T_{\rm aux}=\infty$ in the auxiliary leads \cite{Kennes12}, 
used also for the other results presented here, as well as with 
$T_{\rm aux}=0$ \cite{Karrasch10a}. We find excellent agreement. We expect that 
the precise value of the few percent optimal 
efficiency gain and its position are weakly affected by higher order 
terms not captured by us. 

\section{Summary} 
We have discussed the charge and heat transport properties of an
elementary single-level quantum dot `heat engine' which can be used to convert waste heat 
into electrical energy. We have studied the linear response and 
nonequilibrium steady-state thermoelectric properties in the absence of local Coulomb 
correlations.
The efficiency and output power of this device was investigated in all details. 
We next included the local two-particle interaction which in meso- or nanoscopic devices 
is an important energy scale. Using a flexible approximate many-body method which can be applied 
in linear response as well as the in the nonequilibrium steady state we were able to 
present a comprehensive picture of the correlation effects on the efficiency and output power.

\acknowledgments{We are grateful to S.~Andergassen, C.~Van den Broeck, and M.~Laakso 
for enlightening discussions. This
work was supported by the DFG via FOR 723 (DK and VM) as well as the Emmy-Noether program (DS).}



\begin{thebibliography}{99}
\bibitem{footnote} We are only interested in the part of the parameter space in which 
our  setup can be used to gain electrical energy. We thus define $\eta$ and $P$ 
such that both are positive under this condition (absolute 
values of the currents).
\bibitem{footnoteLB} For $U \neq 0$ the Landauer-B\"uttiker formulas 
for the currents can in general not be used anymore. One has to resort 
to the more general approach of Ref.~\cite{Meir92}. However, in our lowest order 
truncation the expressions for the currents within the latter reduce 
(without any additional approximations) to those of the Landauer-B\"uttiker 
approach with renormalized single-particle parameters \cite{Karrasch10a,Kennes12}.
\end{thebibliography}
\end{document}